\documentclass[journal,article,pdftex,10pt,a4paper,symmetry]{mdpi} 

\usepackage{array,multirow}

\def\al{\alpha}
\def\be{\beta}
\def\ga{\gamma}
\def\de{\delta}
\def\ep{\epsilon}
\def\ve{\varepsilon}

\def\et{\eta}
\def\th{\theta}

\def\ka{\kappa}
\def\la{\lambda}

\def\rh{\rho}

\def\si{\sigma}

\def\ps{\psi}

\def\Ga{\Gamma}

\def\cF{{\cal F}}

\def\cL{{\cal L}}
\def\cM{{\cal M}}

\def\mn{{\mu\nu}}
\def\kl{{\ka\la}}

\def\rs{{\rh\si}}

\def\fr#1#2{{{#1} \over {#2}}}
\def\half{{\textstyle{1\over 2}}}
\def\quar{{\textstyle{1\over 4}}}

\def\frac#1#2{{\textstyle{{#1}\over {#2}}}}

\def\abs#1{\left|{#1}\right|}

\def\lsim{\mathrel{\rlap{\lower4pt\hbox{\hskip1pt$\sim$}}
    \raise1pt\hbox{$<$}}}
\def\gsim{\mathrel{\rlap{\lower4pt\hbox{\hskip1pt$\sim$}}
    \raise1pt\hbox{$>$}}}
\def\sqr#1#2{{\vcenter{\vbox{\hrule height.#2pt
         \hbox{\vrule width.#2pt height#1pt \kern#1pt
         \vrule width.#2pt}
         \hrule height.#2pt}}}}

\def\prt{\partial}

\def\etal{{\it et al.}}

\def\pt#1{\phantom{#1}}

\def\ol#1{\overline{#1}}

\def\vb#1#2{e_{#1}^{{\pt{#1}}#2}}
\def\ivb#1#2{e^{#1}_{{\pt{#1}}#2}}
\def\uvb#1#2{e^{#1#2}}

\def\ab{\overline{a}{}}

\def\cb{\overline{c}{}}

\def\sb{\overline{s}{}}

\def\twiddle{\lower4pt\hbox{\hskip-0pt{$\widetilde{}$}}}
\def\m@th{\mathsurround=0pt}
\def\cmapstochar{\mathrel{\rlap{
  \lower0.1pt\hbox{\hskip-1.75pt{$\mapstochar$}}}
  \raise0pt\hbox{\hskip2.5pt{$\twiddle$}}}}
\def\notsimfill{$\m@th\cmapstochar$}
\def\scroodle#1{\vbox{\ialign{##\crcr\notsimfill\crcr
  \noalign{\kern-4pt\nointerlineskip}
   $\hfil\displaystyle{#1}\hfil$\crcr}}}

\def\cmapstocharbig{\mathrel{\rlap{
  \lower0.1pt\hbox{\hskip0.25pt{$\mapstochar$}}}
  \raise0pt\hbox{\hskip4.5pt{$\twiddle$}}}}
\def\notsimfillbig{$\m@th\cmapstocharbig$}
\def\scroodlebig#1{\vbox{\ialign{##\crcr\notsimfillbig\crcr
  \noalign{\kern-4pt\nointerlineskip}
   $\hfil\displaystyle{#1}\hfil$\crcr}}}

\def\mt{m^{\rm T}}
\def\ms{m^{\rm S}}

\def\af{(a_{\rm{eff}})}

\def\afb{(\ab_{\rm{eff}})}

\def\abt{(\ab^{\rm T}_{\rm{eff}})}
\def\abs{(\ab^{\rm S}_{\rm{eff}})}

\def\cbt{(\cb^{\rm T})}
\def\cbs{(\cb^{\rm S})}

\def\ctw{\scroodle{c}{}}

\def\lrpartial{\raise 1pt\hbox{$\stackrel\leftrightarrow\partial$}}

\def\lrDmu{\stackrel{\leftrightarrow}{D_\mu}}

\def\G{G_N}

\newcommand{\beq}{\begin{equation}}
\newcommand{\eeq}{\end{equation}}
\newcommand{\bea}{\begin{eqnarray}}
\newcommand{\eea}{\end{eqnarray}}
\newcommand{\bit}{\begin{itemize}}
\newcommand{\eit}{\end{itemize}}
\newcommand{\rf}[1]{(\ref{#1})}

\def\sbh{\hat{\sb}{}}

\def\epd{s^{(d)}}

\def\C{\v{C}}

\def\G{G_N}

\def\pp{|\vec p|}
\def\ml{|\vec l|}
\def\Fd{\cF^w(d)}
\def\Fdw#1{\cF^{#1}(d)}

\def\mw{m_w}

\def\re{{\rm Re}~}
\def\im{{\rm Im}~}

\def\cof#1#2{{\ol s}^{(#1)}_{#2}{}}

\def\K{\mathcal K}

\def\Kd{{\K}^{(d)}{}}
\def\KHat{\widehat{\K}^{(d)}{}}

\def\cHat{\widehat{s}}
\def\bHat{\widehat{q}}
\def\dHat{\widehat{k}}

\def\ct{\overline s}
\def\ctHat{\skew{3}\widehat{\ct}}

\def\dc{\circ}

\firstpage{1} 
\articlenumber{x}
\doinum{10.3390/------}
\pubvolume{xx}
\pubyear{2016}
\copyrightyear{2016}
\externaleditor{}
\history{Received: date; Accepted: date; Published: date}
\pdfoutput=1

\Title{The Standard-Model Extension and Gravitational Tests}

\Author{Jay D.\ Tasson}
\AuthorNames{Firstname Lastname, Firstname Lastname and Firstname Lastname}

\address[1]{
Department of Physics and Astronomy,
Carleton College, One North College St.\
Northfield, MN 55057 USA; jtasson@carleton.edu}

\abstract{
The Standard-Model Extension (SME) provides a comprehensive effective field-theory
framework for the study of CPT and Lorentz symmetry. This work reviews the structure and
philosophy of the SME and provides some intuitive examples of symmetry violation. The results of
recent gravitational tests performed within the SME are summarized including analysis of results
from the Laser Interferometer Gravitational-Wave Observatory (LIGO), sensitivities achieved in
short-range gravity experiments, constraints from cosmic-ray data, and results achieved by studying
planetary ephemerids. Some proposals and ongoing efforts will also be considered including
gravimeter tests, tests of the Weak Equivalence Principle, and antimatter experiments. Our review
of the above topics is augmented by several original extensions of the relevant work. We present
new examples of symmetry violation in the SME and use the cosmic-ray analysis to place first-ever
constraints on 81 additional operators.}

\keyword{gravity, Lorentz symmetry, CPT symmetry}

\begin{document}



\section{Introduction}
\label{intro}

Our present description of nature is based on 2 enormously successful theories:
General Relativity (GR),
a classical theory describing all gravitational phenomena,
and the Standard Model (SM) of particle physics,
which provides a quantum description of all other interactions.
It is widely expected that these theories
are merely the low-energy limit of some more fundamental theory
that would take over as the characteristic energies involved in experiments
approach the Planck scale, $10^{19}$ GeV.
Experimental information to guide the development of a Planck-scale theory
would,
by conventional thinking,
come from Planck-energy experiments,
which are likely to remain infeasible far into the future.
An alternative approach is to search for small deviations from known physics
(the SM and GR)
in present-day experiments,
with the hope that small deviations,
if found,
would encode information about the underlying theory.

Lorentz symmetry,
the idea that physical results are unchanged under rotations and boosts of the system,
and CPT symmetry,
the associated invariance of the system under the combination of discrete symmetries
of charge conjugation,
parity,
and time reversal,
are pillars of both the SM and GR.
Hence
violations of these symmetries,
if found,
would provide a novel signal of new physics.
Moreover,
the possibility of violations of these symmetries has been demonstrated
in candidates for the underlying theory,
like strings \cite{ks,kp}. 

The systematic search for Lorentz and CPT violation
using the comprehensive effective field theory based framework
of the Standard-Model Extension (SME) \cite{sme,sme1,akgrav,sme2,sme3,sme4} provides
a method of searching for Planck-suppressed effects in known physics
in a complete and organized way.
In a nut shell,
the SME adds to known physics
all Lorentz and CPT violating effects at the level of the action.
The terms added to the action of the SM and GR to form the SME
are generated from Lorentz and CPT violating operators acting on SM and GR fields
along with coefficients for Lorentz and CPT violation that parameterize
the amount of symmetry violation in the theory.
The addition of Lorentz and CPT violating terms can be thought of as a series expansion
about known physics in ever increasing mass dimension of the operators involved.
The SME coefficients can then be sought in experiment.
Over 1000 limits on SME coefficients have been set via experiment and observation \cite{data},
but much remains to be explored,
particularly in the case of the so-called nonminimal operators
of mass dimension greater than 4,
where few constraints have been set by the direct analysis of experimental data.
It should be emphasized that the SME is a test framework
designed for a broad search for yet-unobserved symmetry violation,
a philosophy that is quite different from model building.
Though the SME is unique in providing a comprehensive
test framework at the level of the action \cite{review},
other approaches to the study of Lorentz and CPT violation exist \cite{oreview}
and the idea of a general test framework
over specific models has philosophical resonance with efforts to parameterize
deviations from GR \cite{ppn,ppe}.

In the next section,
Sec. \ref{basics},
a basic introduction
to both Lorentz and CPT symmetries
is provided along with introductory-physics level examples 
that illustrate 
behaviors that arise from violations of these symmetries.
Following these basics,
Sec.\ \ref{SME} provides a summary of the SME philosophy
and the effective-field theory based structure it employs.
An alternative and somewhat more in-depth treatment
paralleling Sec.\ \ref{intro} - \ref{SME}
can be found in Ref.\ \cite{review}
along with some more general review of other areas of activity within the SME.
Section \ref{grev}
reviews recent experimental results and phenomenological proposals 
that are connected with gravitational physics.
One recent work \cite{cer} places tight initial constraints 
on 74 Lorentz-violating operators of mass dimension 6 and 8 in the pure-gravity sector.
In Sec.\ \ref{cer}
we extend this analysis to obtain another 81 
tight initial constraints on dimension 10 operators.
Throughout this work we use natural units
and metric signature $+2$.

\section{Symmetry Violation}
\label{basics}

In this section we consider several examples of symmetries and their violation.
Though the examples are comparatively simple,
the nature of these examples
will map directly onto the SME structure in Sec.\ \ref{SME}.
Lorentz symmetry contains both rotations and boosts.
In the first subsection we will appeal to the visual nature of rotation invariance
and consider examples of rotation invariance and rotation-invariance violation
as examples of the Lorentz-symmetry case.
In the second subsection, CPT symmetry and CPT violation will be considered.

\subsection{Rotations}

To begin our example,
consider the classical, nonrelativistic Lagrangian for a particle of mass $m$,
position $\vec r$, and charge $q$ in a magnetic field $\vec B$ given by a vector potential $\vec A$
($\vec B = \vec \nabla \times \vec A$):
\beq
L = \half m \dot{\vec r}^2  + q \dot {\vec r} \cdot  \vec A.
\label{la}
\eeq
We will first do what is known as an {\it observer} transformation on this system,
in this case a rotation.
This corresponds to the experimenter turning his or her head
and taking their coordinates with them.
This transformation is carried out
by acting with the standard rotation matrix $\mathbf R$
on all vector components such that the components of a generic vector $\vec V$
transforms as $V_j \rightarrow \vec V_{j^\prime} = R_{j^\prime j} V_j$.
Doing this transformation
to all vector components in \rf{la} reveals that it is form invariant.
This is a signal that the theory in \rf{la} possesses ``observer-rotation invariance''
as might have been expected.
In other words,
the outcome of experiments governed by \rf{la} does not depend on the coordinates used.

To see even more explicitly that the theory is invariant
we can perform the following series of steps:
(i) set up a system with some initial conditions,
(ii) calculate the final configuration using
the original theory, where in the example above,
the acceleration of the particle $\vec a$ suffices
as a proxy for the final configuration,
(iii) apply the symmetry transformation to the result,
(iv) apply the symmetry transformation to the initial conditions
(v) calculate the final configuration
using the transformed theory.
As shown in figure 1,
when these steps are applied for observer rotations
the results of steps (iii) and (v) match, reflecting
the obvious observer-rotation invariance of the system.
\begin{figure}[H]
\centering
\includegraphics[width=13cm]{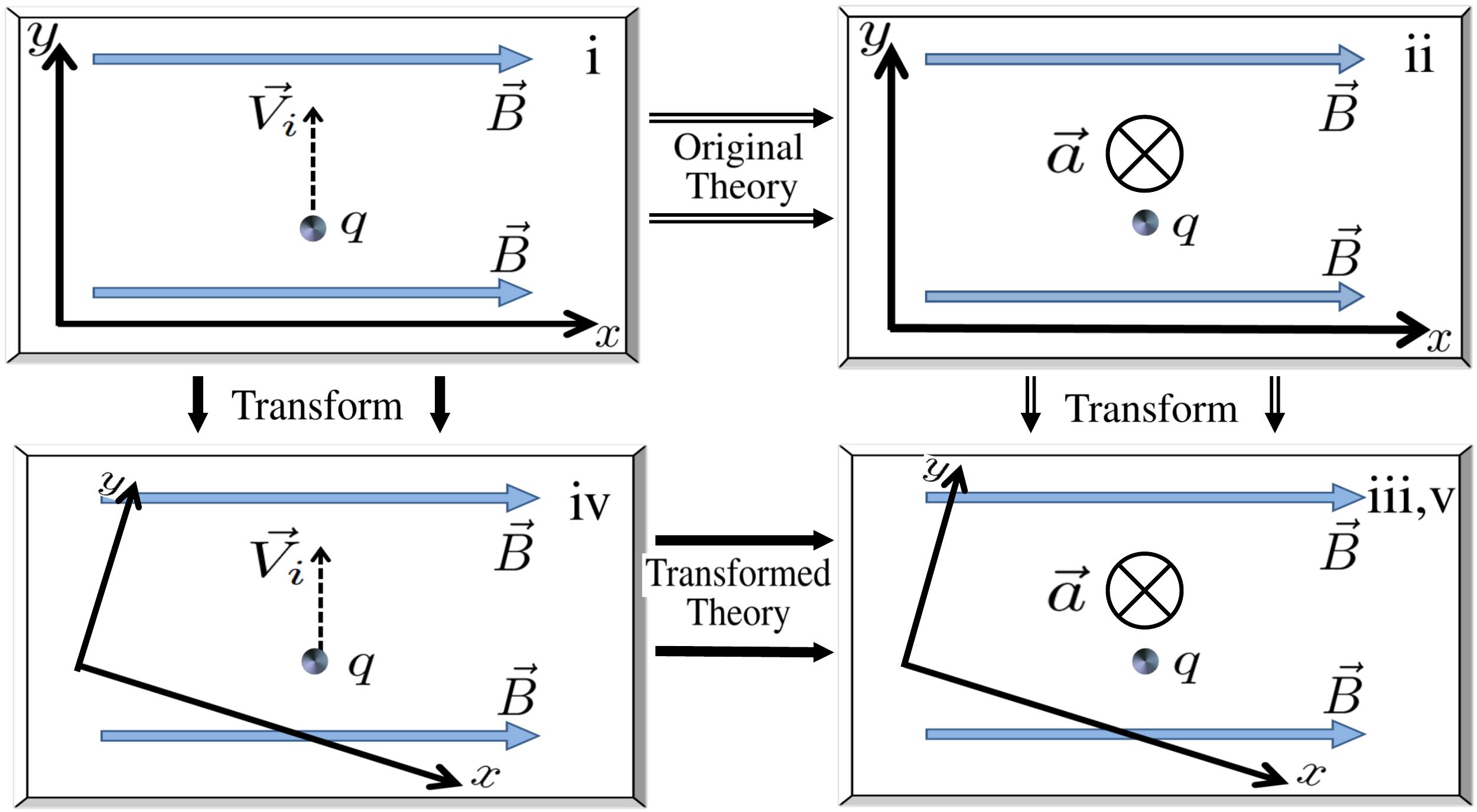}
\caption{Illustration of observer symmetry in a sample system.}
\end{figure} 

We can next apply the same procedure for a {\it particle} rotation.
This procedure leaves the observer and the coordinates fixed
but rotates all fields.
Operationally,
the procedure is carried out the same way on this rotation-invariant system,
and as result of the symmetry the outcome will be identical.
While we could draw this procedure in an identical way,
observer and particle transformations will be distinct when spacetime symmetries are broken
and we draw the pictures in Fig.\ 2 for the particle transformation case somewhat differently 
to highlight the difference
between rotating the coordinates and rotating the physical system.
\begin{figure}[H]
\centering
\includegraphics[width=13cm]{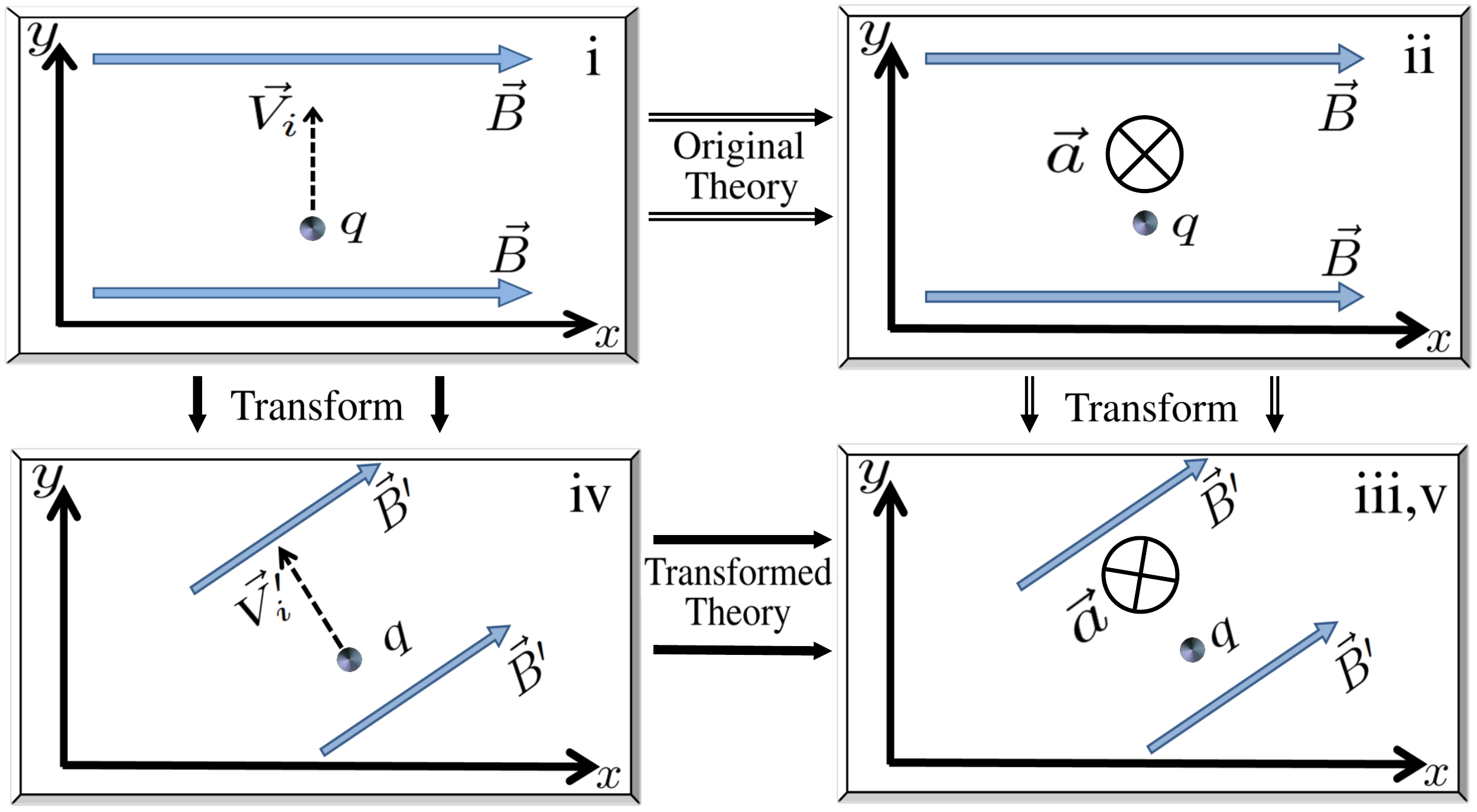}
\caption{Illustration of particle-rotation symmetry in a sample system.}
\end{figure} 

We can now use this system as a toy model for broken Lorentz invariance.
Suppose that the experimenter were unaware of the physics generating the 
vector potential and hence the magnetic field.
Perhaps it exists on much larger scale than their lab.
If they now perform a particle rotation on ``the system''
they will rotate the apparatus in their lab,
but not the magnetic field.
Here the transformed theory
will be
\bea
L &=& \half m \dot{\vec r}^{\prime 2}  + q \dot{\vec r}^\prime \cdot  \vec A
\nonumber \\
  &=& \half m \dot{\vec r}^2  + q \dot{\vec r} \mathbf R^T \cdot  \vec A,
\label{la2}
\eea
where $\mathbf R^T$ is the transposed rotation matrix.
Note that the theory is no longer particle-rotation invariant
as the form has changed.
Applying this transformation
to our cartoon example
leads to the situation shown in Fig.\  3
in which the acceleration found in the transformed system
is different (acceleration is smaller under the conditions shown)
than the original system.
\begin{figure}[H]
\centering
\includegraphics[width=13cm]{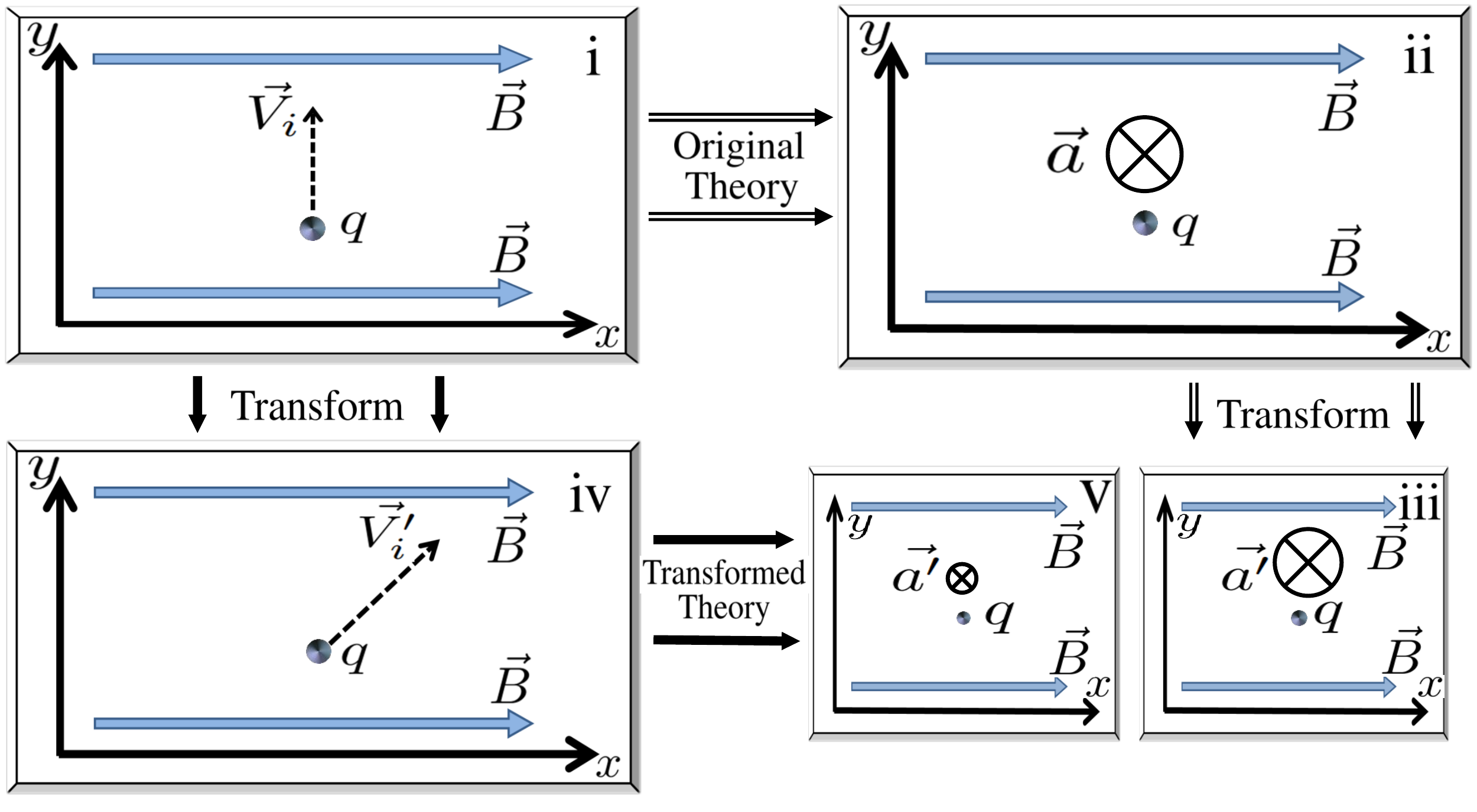}
\caption{Illustration of effective particle-rotation symmetry violation in a sample system.}
\end{figure}
Workers in the lab will then know if the system is particle-Lorentz invariant or not
by performing their experiment,
then rotating it,
then comparing the results.
If the acceleration is different in the rotated system,
particle-Lorentz invariance is broken.
In our discussion of the SME to follow,
it will be large-scale fields
called coefficients for Lorentz violation,
perhaps associated with spontaneous breaking of Lorentz symmetry at the Planck scale,
that will play the role of $\vec B$ and be sought in this manner.
We also note in passing
that if an undetected large-scale ``conventional'' field existed in the lab,
it could also be detected in this way,
an idea that has been applied to efforts to detect spacetime torsion \cite{torsion}
and gravitomagnetic effects in the lab \cite{jtmag}.
Additional examples similar to the one presented here
can be found in Refs.\ \cite{review,jttb}.

\subsection{CPT}

The role of CPT symmetry
and CPT violation can also be illustrated in the above example.
Under CPT,
$q$ and $\vec A$ change signs,
but nothing else.
Hence the theory is invariant.
A subtlety is that although $\vec A$ changes sign,
$\vec B$ does not.
Figure 4 illustrates the usual procedure for seeing that our example
is CPT invariant.
\begin{figure}[H]
\centering
\includegraphics[width=13cm]{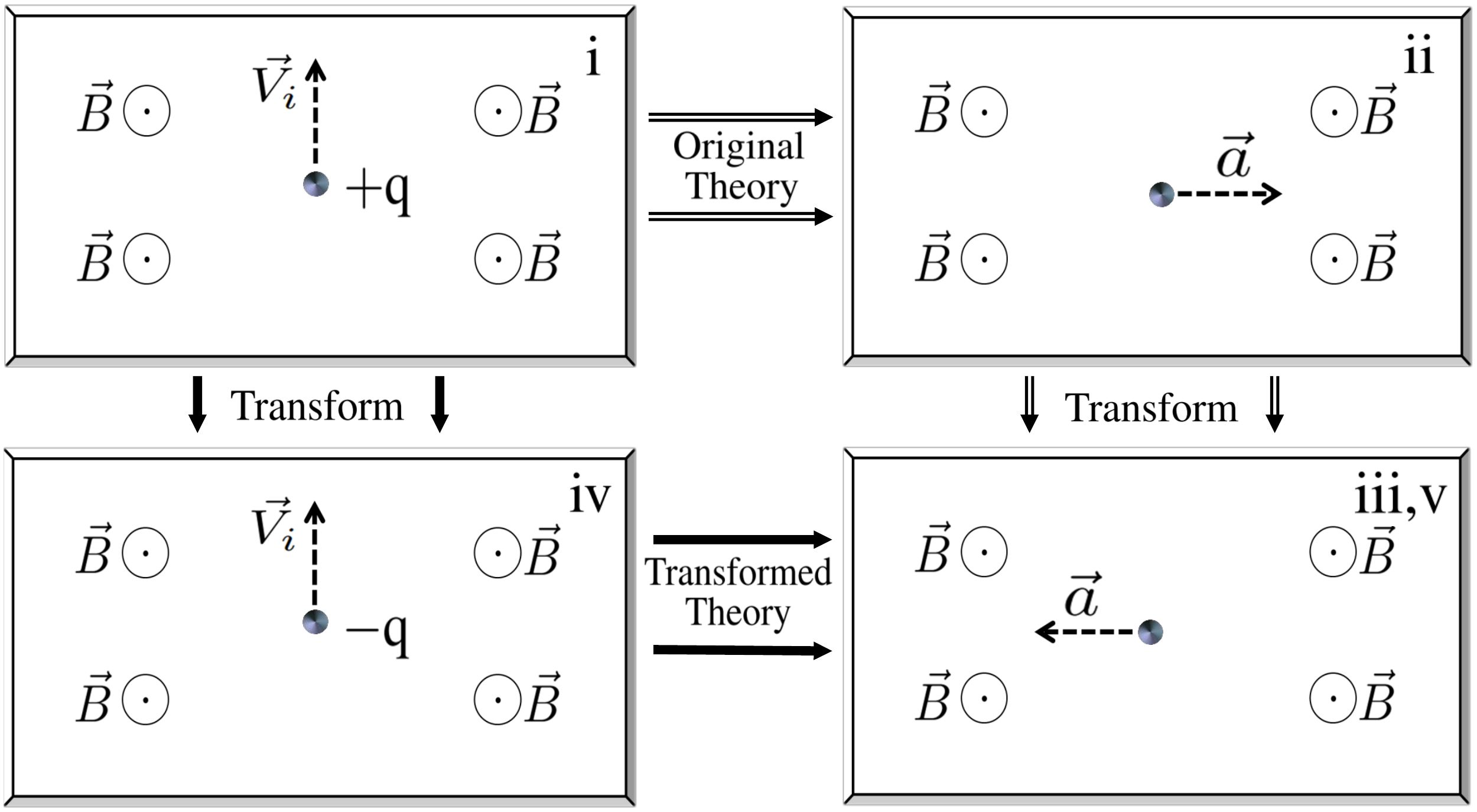}
\caption{Illustration of CPT symmetry in a sample system.}
\end{figure}

If we again treat $\vec A$ as a nontransforming background
while workers in the lab build the CPT transformed version of their device,
then $\vec A$ will not change under CPT
and the second term in the theory will change sign
for the laboratory CPT transformed system.
Hence the presence of a background $\vec A$ leads to effective CPT violation
as well as Lorentz violation.
The effective CPT violation induced by $\vec A$ is illustrated in Fig.\ 5.
\begin{figure}[H]
\centering
\includegraphics[width=13cm]{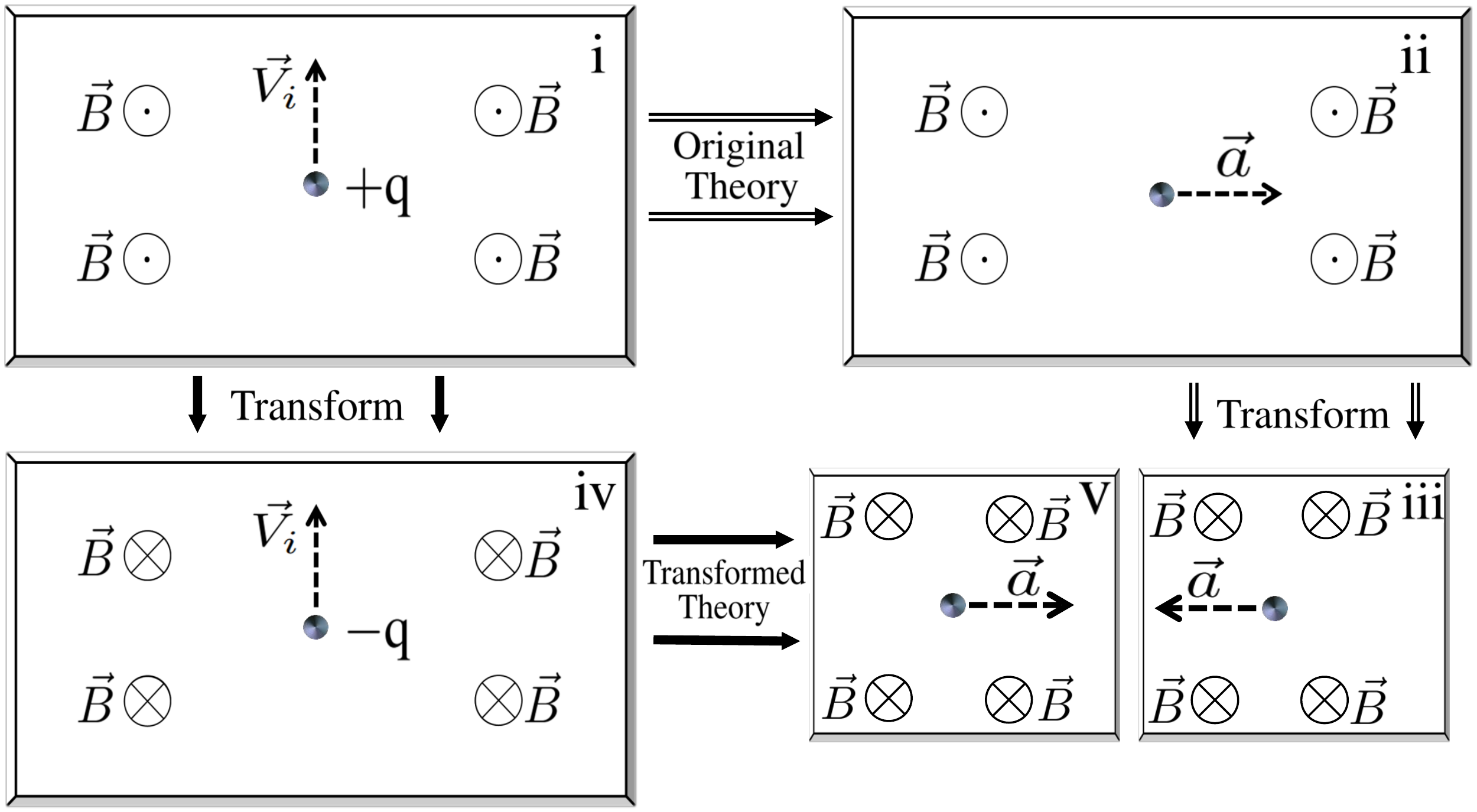}
\caption{Illustration of effective CPT-symmetry violation in a sample system.}
\end{figure}

The violation of both CPT and Lorentz symmetry is a feature shared quite generally by backgrounds
with odd numbers of indices.
For a somewhat more technical review of the connection between CPT and Lorentz symmety,
see the contribution to this issue by Lehnert \cite{rlsym}.
For simple examples having backgrounds with even numbers of indices,
see Refs.\ \cite{review,jttb}.

\section{The SME}
\label{SME}

In this section we present the SME action with focus on the sectors relevant
for the discussion and development of the gravitational results considered in the rest of this work.
For a similar review focused on neutrinos,
see the contribution to this issue by D\'iaz \cite{jdiaz}.
For some review of other sectors,
see Ref.\ \cite{review}.
Gravitational phenomenology in the SME can be thought of as originating in 2 places:
the pure-gravity sector, which describes the dynamics of the gravitational field itself,
and gravitational couplings in the other sectors.
We consider first the gravitationally coupled fermion section,
then the pure-gravity sector below.

\subsection{Gravitationally coupled matter}

The implications of gravitational couplings in the 
minimal fermion sector have been studied considerably.
Here we review the structure of this limit,
and use it as an example of the SME structure more generally.
The Lagrange density for the minimal fermion sector with gravitational couplings
reads \cite{akgrav}
\beq
\cL_{\ps-g} = 
\half i e \ivb \mu a \ol \ps \Ga^a \lrDmu \ps 
- e \ol \ps M \ps,
\label{fermion}
\eeq
where
\bea
\Ga^a
&\equiv & 
\ga^a - c_{\mu\nu} \uvb \nu a \ivb \mu b \ga^b
- d_{\mu\nu} \uvb \nu a \ivb \mu b \ga_5 \ga^b
- e_\mu \uvb \mu a 
- i f_\mu \uvb \mu a \ga_5 
- \half g_{\la\mu\nu} \uvb \nu a \ivb \la b \ivb \mu c \si^{bc}, \\
M
&\equiv &
m + a_\mu \ivb \mu a \ga^a 
+ b_\mu \ivb \mu a \ga_5 \ga^a 
+ \half H_{\mu\nu} \ivb \mu a \ivb \nu b \si^{ab}.
\eea
Here $\ps$ is the fermion field,
$D_\mu$ is the covariant derivative for gravity as well as $U(1)$,
and
$\vb \mu a$ is the vierbein,
which, along with its determinant $e$ and the covariant derivative,
provides the couplings to gravity
by linking each point on spacetime manifold with a Minkowski tangent space.
The Dirac gamma matricies are denoted $\ga^a$,
and $\ga_5$ and $\si^{ab}$ denote standard combinations there of \cite{iz}.
The objects $a \ldots H$ are coefficient fields for Lorentz violation,
which in general are different for different types of particles.
The interpretation of the coefficient fields is developed further below.
The field-theoretic and geometric structure suggested by the objects
introduced above is rich,
but a full review of this would take us too far afield
from the phenomenological development which is the primary subject of this work.
For more on these issues,
see Ref.\ \cite{review} and references there in.

Before developing the gravitational implications of \rf{fermion},
we point out several features of \rf{fermion} 
with the goal of generating a broader understanding
of the SME structure.
First,
note that the form of the Minkowski-spacetime SME \cite{sme1} can be recovered
in the limit $\vb \mu a \rightarrow \de^a_\mu$,
and the standard Lorentz-invariant Dirac Lagrange density can be recovered
with the additional restriction $a \ldots H \rightarrow 0$.
This last limit reflects a key feature of the SME structure:
it contains known physics and can be thought of as 
providing an expansion about known physics in Lorentz-violating operators.
The action \rf{fermion} is known as the minimal SME limit
as it contains operators of dimension 3 and 4
that are power-counting renormalizable.
For a basic introduction to the idea of operator dimension,
see Ref.\ \cite{leap16}.
The expansion about known physics can be continued beyond dimension 3 and 4
and the full series of Lorentz violating operators in the fermion sector has been studied \cite{sme4}
in the Minkowski-spacetime limit, as have other sectors \cite{sme2,sme3}.
Consideration of higher dimension operators in the pure-gravity sector has also begun
\cite{xu,cer,kmgw,qbho}
and the key points are reviewed in the next subsection.

The content of the coefficient fields $a \ldots H$
is in principle determined by how Lorentz symmetry is broken,
though in practice we proceed under general considerations
without specializing to a particular model of Lorentz-symmetry breaking.
The ways in which Lorentz-symmetry can be broken
can be divided into 2 classes,
explicit and spontaneous.
Explicit breaking involves coefficients for Lorentz violation
that are externally prescribed properties of the spacetime.
When Lorentz violation arises spontaneously,
the underlying theory is Lorentz-invariant,
but the low-energy solutions violate Lorentz symmetry
through a process of spontaneous symmetry-breaking analogous
to electroweak symmetry breaking in the SM.
In flat-spacetime studies,
the coefficient fields are typically taken as constant coefficients,
satisfying, for example, $\prt_\al c_\mn = 0$.
This can be thought of as either a particular explicit choice
or as a search for a constant vacuum expectation value given to the field
via spontaneous breaking.
The consideration of constant coefficients maintains energy-momentum conservation
and could be thought of as a leading contribution to more complex spacetime dependence.
In curved spacetime,
constant coefficients fields are typically not compatible with
a theory of gravity based on Riemannian geometry \cite{akgrav,rb1,rb2}
and specialization to spontaneous breaking is chosen
for most SME phenomenology.
Here the coefficient fields can be thought of as involving a constant vacuum value
plus a fluctuation about the vacuum, for example
$c_\mn = \cb_\mn + \ctw_\mn$.
In the asymptotically flat studies considered here,
the vacuum values such as $\cb_\mn$ are taken as constant 
and can be identified with the coefficients for Lorentz violation
explored in flat-spacetime studies.
Hence we refer to the vacuum values as the coefficients for Lorentz violation.
Under fairly general assumptions 
in the work considered here,
the fluctuations can be integrated out
prior to the development of phenomenology \cite{lvgap,akjt,lvpn}.
An approach that considers non-constant coefficient fields
under a different set of assumptions
has also now been developed \cite{chuck}.

For most gravitational experiments,
the classical, post-Newtonian implications
of the relativistic quantum field theory \rf{fermion}
is what's relevant.
As a result,
tools corresponding to \rf{fermion}
including the relativistic hamiltonian,
the nonrelativistic hamiltonian,
and the classical Lagrangian
have been developed
to 3rd post newtonian order \cite{lvgap}.
Note that the matter theory impacts the structure of the spacetime
since the metric is derived from the matter theory
involving coefficients for Lorentz violation associated with the source,
as well as the motion of test particles in that spacetime,
which involves coefficients for Lorentz violation associated with the test particle.
As an example of the tools that result,
in the newtonian limit
the equation of motion for a particle in an Earth-based laboratory
associated with spin-independent effects takes the form
\beq
F_j = m_{jk} \ddot{x}_k,
\label{labeom}
\eeq
where
\beq
m_{jk} = 
\mt \left(1 + \cbt_{tt} \right) \de_{jk} 
+ 2 \mt \cbt_{(jk)},
\label{inertialm}
\eeq
and the vertical component of the force is
\beq
F_z = 
- \mt g \Big[1 
+ \fr{2 \al}{\mt} \abt_t
+ \fr{2 \al}{\ms} \abs_t 
+ \cbt_{tt} + \cbs_{tt} \Big].
\eeq
Here $\al$ is a constant that characterizes couplings in the underlying theory
of spontaneous breaking,
$\mt$ and $\ms$ are masses of the test and source bodies respectively,
and superscripts T and S refer to the coefficients associated with the
test and source bodies respectively.
The combindation $a_\mu - m e_\mu$ is denoted $\af_\mu$.
Note that in addition to generating the annual and sidereal variations associated
with boost and rotation invariance in the lab,
violations of the Weak Equivalence Principle (WEP)
are induced by the particle-species dependence in this result.

\subsection{Pure gravity}

Maxwell's electrodynamics is famously a linear theory,
meaning that if one takes the potential $A_\mu$ as the fundamental object,
the field equations are linear in $A_\mu$.
This permits linear superposition as a convenient method of constructing solutions.
General Relativity
is a nonlinear theory of gravity based on the curvature of spacetime
encoded in the spacetime metric $g_\mn$,
which can be written in terms of the Minkowski metric $\et_\mn$ plus an object $h_\mn$,
typically called the metric perturbation.
By analogy with electrodynamics,
if $h_\mn$ is taken as the fundamental object,
the GR field equations are nonlinear in $h_\mn$,
hence nonlinear Lorentz-violating corrections to GR might be expected in an SME expansion
about the GR action.
However it should also be noted that such nonlinearities are negligibly small in practice
in large classes of experiments
where the spacetime can be considered asymptotically Minkowski
and $h_\mn$ can be treated as a small correction.

Progress toward phenomenology in the pure-gravity sector of the SME
began with the development \cite{akgrav} and exploration \cite{lvpn} of its minimal limit.
While the construction at the level of the action developed in these works
includes the full nonlinear theory of gravity,
the development of phenomenological tools is specialized to the linearized-gravity limit \cite{lvpn}.
More recently,
exploration of the nonminimal pure-gravity sector has begun \cite{xu,cer,kmgw,qbho},
including consideration of some terms associated with the nonlinearities of gravity \cite{qbho}.
To provide a maximally coherent treatment of phenomenology and experiment,
we consider first the full SME expansion under the restriction of linearized gravity,
then offer some comments about current proposals to explore effects in nonlinear gravity.

The generic form of the Lagrange density in the linearized limit
including both the Lorentz-violating and Lorentz-invariant contributions can be written \cite{kmgw},
\beq
\cL_{\Kd} = \quar h_\mn \KHat^{\mn\rs} h_\rs ,
\label{L1}
\eeq
where the operator
\beq
\KHat^{\mn\rs}=
\Kd^{\mn\rs\ve_1\ve_2\ldots\ve_{d-2}}
\prt_{\ve_1} \prt_{\ve_2}\ldots \prt_{\ve_{d-2}}
\label{KHat}
\eeq
has mass dimension $d \geq 2$,
and the coefficients $\Kd^{\mn\rs\ve_1\ve_2\ldots\ve_{d-2}}$ are taken as constant and small.
An exploration of the operator \rf{KHat}
by decomposition into irreducible parts reveals 14 classes of operators are involved.
Of these,
3 classes, written as follows,
\beq
\KHat^{\mn\rs} \supset \cHat^{\mn\rs} + \bHat^{\mn\rs} + \dHat^{\mn\rs}, 
\label{lingrav1}
\eeq
are consistent with the usual gauge invariance of GR.
The term 
\beq
\cHat^{\mn\rs} = \sum_{{\rm even}\,\, d \geq 4} s^{(d)}{}^{\mu\rh\dc\nu\si\dc^{d-3}}
\eeq 
contains operators at each even dimension greater than or equal to 4,
hence it contains a minimal contribution.
Here a circle denotes a contraction with a partial derivative. 
The contribution
\beq
\bHat^{\mn\rs} =  \sum_{{\rm odd}\,\, d \geq 5} q^{(d)}{}^{\mu\rh\dc\nu\dc\si\dc^{d-4}},
\eeq
involves operators of odd dimension greater than or equal to 5,
and 
\beq
\dHat^{\mn\rs}= \sum_{{\rm even}\,\, d \geq 6} k^{(d)}{}^{\mu\dc\nu\dc\rh\dc\si\dc^{d-5}}, 
\eeq
contributes operators of even dimension greater than or equal to 6. 
For further discussion of their properties,
see \cite{kmgw}.
Study to date has been focused on these operators,
which are associated with spontaneous Lorentz violation.

Operators $\bHat^{\mn\rs}$ and $\dHat^{\mn\rs}$ are associated with birefringence of gravitational waves.
Constraints on the dimension 5 and 6 coefficients they contain \cite{kmgw} 
have been placed using the initial direct observation
of gravitational waves by LIGO \cite{ligo}.
Constraints based on the dispersion triggered by isotropic combinations
of coefficients at dimentions 4, 5, and 6 have also been attained \cite{yypgw}
based on LIGO observations of merger events \cite{ligo,ligo2}.
We note in passing that LIGO has also been used to achieve photon-sector sensitivities
by interpreting a suitable aspect of the 2006-2007 ligo data 
as providing a Michelson-Morley-type test \cite{kmgw2}.
Matter-sector implications have also been noted \cite{msgw}.

The nature of $\cHat^{\mn\rs}$ can be elucidated by introducing a dual operator as follows:
$\cHat{}^{\mu\rh\nu\si}
=-\ep^{\mu\rh\al\ka}\ep^{\al\nu\si\be\la}\ctHat{}_\kl \prt_\al \prt_\be$.
In term of the dual,
the Lagrange density can be written,
\beq
\cL = \quar \ep^{\mu\rh\al\ka} \ep^{\nu\si\be\la} 
h_\mn (\et_\kl - \ctHat{}_\kl) \prt_\al\prt_\be h_\rs, + 
\quar h_\mn 
(\bHat{}^{\mu\rh\nu\si} 
+ \dHat{}^{\mu\nu\rh\si})
h_\rs,
\eeq
where $\et_\kl$ in the first term is the appropriate limit of the standard Einstein-Hilbert contribution.
The use of the dual operator reveals that $\cHat{}^{\mu\rh\nu\si}$ acts as a momentum-dependent
perturbation to the Minkowski metric.
As in the case of the general operator
introduced in Eq.\ \rf{lingrav1},
$\sbh_\mn$ can be expanded in terms of coefficients for Lorentz violation
at each dimension $d$,
which are taken as constant and small in most studies,
and an appropriate number of derivatives.
Explicitly,
\beq
\sbh_\mn \equiv 
\sum_{{\rm even} \,\, d \geq 4} 
(\sb^{(d)})_{\mn}{}^{\al_1 \ldots \al_{d-4}} 
\prt_{\al_1} \ldots \prt_{\al_{d-4}}.
\label{sdef}
\eeq
Here $\sb^{(4)}_\mn \equiv \sb_\mn$ is the coefficient studied
in numerous minimal SME investigations as noted in the next section.
An additional coefficient coupling to the Weyl tensor, $t^{\mn \rh \si}$ was considered
in the original investigation of the minimal gravity sector \cite{akgrav},
but phenomenological implications stemming from this coefficient have not been found \cite{ybt}.
Additionally, elements of the sum \rf{sdef} 
at dimension 4,6, and 8 have been sought via gravitational \v Cerenkov radiation.
We summarize this method and expand on the result in Sec.\ \ref{cer} of this work
where constraints on the $d=10$ coefficients are achieved.

A combination of $\cHat{}^{\mu\rh\nu\si}$ and $\dHat{}^{\mu\rh\nu\si}$ coefficients at $d=6$
named $(\bar k_{\rm eff})_{jklm}$ 
have been identified \cite{xu} as contributing to short-range gravity experiments via the modified Poisson equation
\beq
- \nabla^2 U = 4 \pi G_N \rh + (\bar k_{\rm eff})_{jklm} \prt_j \prt_k \prt_l \prt_m U.
\eeq
Here $U(\vec r)$ is the modified Newtonian gravitational potential, and
$G_N$ is Newton's constant.

Initial studies of Lorentz violation in nonlinear gravity
have also been performed \cite{qbho}
that focus on the $d=8$ (12 index) coefficient field that couples to 3 Riemann curvature tensors
in the action.
The phenomenology of a subset of the vacuum values 
associated with this coefficient field
denoted $K_{jk}$ has been established.

\begin{table}[H]
\begin{center}
\begin{tabular}{c|c|c|c}
system              & coefficients                & proposal & constraints \\
\hline\hline
gravitational \v Cerenkov radiation & $\sb^{(4)}_\mn$, $\sb^{(6)}_{\mn \al_1 \al_2}$, $\sb^{(8)}_{\mn \al_1 \al_2 \al_3 \al_4}$,
$\sb^{(10)*}_{\mn \al_1 \al_2 \al_3 \al_4 \al_5 \al_6}$
& \cite{cer} & \cite{cer} [$*$]\\
superconducting gravimeters & $\afb_\mu$, $\cb_\mn$, $\sb^{(4)}_\mn$  & \cite{lvpn,lvgap} & \cite{flowers} \\
short-range gravity devices & $(\bar k_{\rm eff})_{jklm}$ $\afb_\mu$ & \cite{xu,lvgap,lvpn} 
& \cite{cshao1,cshao2,long1,panjwani,long2} \\
gravitational-wave interferometers & $\sb^{(4)}_\mn$, $\sb^{(6)}_{\mn \al_1 \al_2}$, $q^{(5)}{}^{\mu\rh\al\nu\be\si\ga}$, $k^{(6)}{}^{\mu\al\nu\be\rh\ga\si\de}$ & \cite{kmgw} & \cite{kmgw, yypgw} \\
lunar laser ranging & $\sb^{(4)}_\mn$ & \cite{lvpn} & \cite{battat,bourgoin} \\
binary-pulsar observations & $\sb^{(4)}_\mn$ & \cite{lvpn} & \cite{lafitte,lshao1,lshao2} \\
planetary ephemerides & $\afb_\mu$, $\sb^{(4)}_\mn$ & \cite{hees} & \cite{hees} \\
gravity probe B & $\sb^{(4)}_\mn$ & \cite{lvpn} & \cite{everett} \\
bound kinetic energy WEP & $\afb_\mu$, $\cb_\mn$ & \cite{lvgap,hm1} & \cite{hm1} \\
atom interferometers & $\afb_\mu$, $\cb_\mn$, $\sb^{(4)}_\mn$ & \cite{lvpn,lvgap} & \cite{hm2,hm3,hm4} \\
comagnetometry & $\afb_\mu$, $\sb^{(4)}_\mn$ & \cite{jtmag} & \cite{jtmag} \\
perihelion precession & $\afb_\mu$, $\cb_\mn$ & \cite{lvgap} & \cite{lvgap} \\
equivalence-principle pendulum & $\afb_\mu$ & \cite{akjt} & \cite{akjt} \\
Solar-spin precession & $\sb^{(4)}_\mn$ & \cite{lvpn} & \cite{lvpn} \\
\end{tabular}
\caption{\label{existing}
Gravitational tests constraining SME coefficients}
\end{center}
\end{table}

\section{Gravitational tests: existing results and proposals}
\label{grev}

A large amount of experimental and observational work has been done 
based on the theory reviewed above,
and numerous proposals exist to extend and improve these results.
In Table 1,
we list systems that have been used to achieve constraints on coefficients for Lorentz violation.
The first column identifies the system;
the second, the coefficients constrained;
the third, references to the associated phenomenological proposals;
and the forth, references to the work in which the constraints were achieved.
Note that in some cases there is overlap between the last two columns.
In some cases
initial constraints on coefficients for Lorentz violation
are achieved in the paper proposing the work
based on a reinterpretation of published information.
In other cases,
the work is essentially proposed in the experimental paper.
An asterisk ($*$) denotes coefficients constrained in
Sec.\ \ref{cer} of this work.
In addition to information provided in the papers cited,
the constraints achieved in each test
are summarized in \cite{data}.

In Table 2, we highlight some of the cases in which existing or planned experiments
could provide improvement over published limits.
Here improvements could involve extending the maximum reach for a given coefficient,
generating sensitivities to new linear combinations of coefficients
that lead to either more independent constraints or a discovery,
or the generation of cleaner constraints that involve fewer assumptions.
The first 3 columns are the same as in Table I.
The last column lists references to some existing and planned experiments associated with the given system
that hold promise for improving sensitivities to the coefficients listed.
Note that no such table can be exhaustive.
The list provided here is merely intended to highlight 
the breath of possibilities.

\begin{table}[H]
\begin{center}
\begin{tabular}{c|c|c|c}
system              & coefficients                & proposal    & experiments\\
\hline\hline
atom interferometer & $\afb_\mu$, $\cb_\mn$, $\sb^{(4)}_\mn$ & \cite{lvgap} & \cite{fai}\\
ring-laser gyroscopes & $\sb^{(4)}_\mn$                  & \cite{nsjt1,nsjt2} & \cite{frlg}\\
torsion pendula & $\afb_\mu$, $\cb_\mn$                  & \cite{lvgap} & \cite{eotwash}\\
binary pulsars & $\afb_\mu$, $\cb_\mn$, $\sb^{(4)}_\mn$ $K_{jk}$      & \cite{qbho,jennings} & \cite{ska}\\
short-range gravity & $K_{jk}$ & \cite{qbho} &  \cite{cshao1,cshao2,long2}\\
gravitational-wave detectors  & $s^{(d)}{}^{\mu\rh\al\nu\si\be \ldots}$, $k^{(d)}{}^{\mu\al\nu\be\rh\ga\si\de \ldots}$, $q^{(d)}{}^{\mu\rh\al\nu\be\si\ga \ldots}$   & \cite{kmgw} & \cite{ligo,ligo2}\\
space-based WEP tests  & $\afb_\mu$, $\cb_\mn$                  & \cite{lvpn,spacewep} & \cite{spacewep, mscope, gg, step}\\
antimatter gravity  & $\afb_\mu$, $\cb_\mn$               & \cite{lvgap,nmipm} & \cite{aegis,alpha,gbar,age}\\
charged matter WEP  & $\afb_\mu$                 & \cite{lvgap} & \cite{charge}\\
muonium free fall & $\afb_\mu$, $\cb_\mn$              & \cite{lvgap} & \cite{muonium}\\
light bending  & $\afb_\mu$, $\cb_\mn$, $\sb^{(4)}_\mn$                  & \cite{qblight} & \cite{lator}\\
time-delay \& Doppler tests & $\sb^{(4)}_\mn$   & \cite{qbtime,lvgap} & \cite{odyssey, astrod, beacon}\\
\end{tabular}
\caption{\label{inprep}
Some systems that could improve upon existing sensitivities.}
\end{center}
\end{table}

\section{Gravitational \C erenkov}
\label{cer}

In this section we obtain and present the first limits
on the dimension 10 coefficients
contained within $\sbh^{\al \be}$
by following the methods of Ref.\ \cite{cer}.
Prior to presenting the results,
we summarizes some of the key ideas used.

As is well known,
when charged particles exceed the phase speed of light in media,
the \v Cerenkov radiation of photons results.
In the presence of Lorentz violation in the SME,
the vacuum exhibits many of the properties of the medium,
and vacuum \v Cerenkov radiation becomes kinematically allowed.
This feature extends to the case of gravitational waves,
where, in the presence of certain coefficients for Lorentz violation,
particles may exceed the speed of gravity and emit \v Cerenkov gravitons \cite{cer,mncer}.
This feature is possible for massive particles as well as photons,
and, in the presence of suitable SME coefficients,
can be both anisotropic and momentum dependent \cite{cer}.

The idea of vacuum gravitational \v Cerenkov radiation can be used to constrain coefficients for Lorentz violation
through consideration of high-energy cosmic rays.
The fact that these rays arrive at Earth from great distances with high energy
limits the amount of energy they could have radiated to gravitational waves,
and hence limits the coefficients for Lorentz violation involved.

To begin our analysis,
we note that the dispersion relation associated with $\sbh^{\al \be}$
can be written
\beq
l_0^2 = \vec l^2 + \sbh^{\al \be} l_\al l_\be,
\label{disp}
\eeq
where $l_\mu$ is the graviton momentum.
The form of Eq.\ \rf{disp} implies that it is convenient to introduce
an effective momentum-dependent vacuum index of refraction for gravity,
\beq
n(\vec l) = \sqrt{1- \sbh^{\al \be} \hat l_\al \hat l_\be}.
\eeq

Standard decay-rate and energy-loss equations can then be used to find
the rate of energy loss to gravitational waves by a faster-than-gravity particle,
which can be written
\beq
\hskip -10pt
\fr {dE}{dt} = 
-\fr{1} {8\pp \sqrt{\mw^2 + \vec p^2}}
\int \fr{d^3l}{ (2 \pi)^2 \ml} |\cM|^2
\de ( \cos \th - \cos \th_C ).
\label{rated}
\eeq
Here $p_\mu$ is the particle momentum,
$m_w$ is its mass,
$\cM$ is a matrix element from quantum field theory,
the angle between $\vec p$ and $\vec l$ is $\th$,
and $\th_C$ is a generalized \v Cerenkov angle that takes the form
\beq
\cos \th_C = 
\fr {\sqrt{\mw^2 + \vec p^2}} {\pp} 
\fr 1 {n(\ml)}
+ \fr{\ml}{2\pp} \left(1-\fr{1}{[n(\ml)]^2}\right).
\label{ca}
\eeq
Inserting the matrix element associated with the given radiating particle
and integrating over graviton momentum yields an explicit form for the power loss to radiated gravitons.
The result is the same for each type of particle considered
(photons, scalars, and fermions) up to a dimension $d$ dependent factor.
This result can then be integrated to get a relation between the initial energy $E_i$
at the start of the particle's trip,
the final energy $E_f$ at detection,
and the time of flight $t$:
\beq
t = \fr {\Fd} {\G (\epd)^2}
\left(\fr{1}{E_f^{2d-5}}-\fr{1}{E_i^{2d-5}}\right).
\label{tflight}
\eeq
Here $\Fd$ is the species-dependent factor.
In the analysis to follow,
we consider fermions for which
\beq
\Fdw \ps = 
\fr {(d-2)(d-3)(2d-3)} {4(2 d^2 - 7d +9)}. 
\label{cdwfermion}
\eeq
The dependence on Lorentz violation in Eq.\ \rf{tflight}
is through the combination
\beq
\epd(\hat p) \equiv (\sb^{(d)})^{\mn\al_1 \ldots \al_{d-4}}
\hat p_\mu \hat p_\nu 
\hat p_{\al_1} \ldots \hat p_{\al_{d-4}},
\eeq
which can also be expressed in terms of spherical harmonics 
and spherical coefficients for Lorentz violation as
\beq
\epd(\hat p) = \sum_{jm} Y_{jm}(\hat p) 
\cof{d}{jm}.
\eeq

Constraints on the coefficients $\epd$ are achieved via high-energy cosmic ray observations from the following projects:
AGASA \cite{cat,agasa},
Fly's Eye \cite{high},
Haverah Park \cite{cat,hp},
HiRes \cite{hires},
Pierre Auger \cite{pa},
SUGAR \cite{cat,sugar},
Telescope Array \cite{ta},
Volcano Ranch \cite{cat,vr},
and
Yakutsk	\cite{cat,yak}.
For details on the events used,
see Table 1 in Ref.\ \cite{cer}.
Extracting a constraint requires knowledge of the initial and final energy
and the distance traveled, $L$.
To generate conservative constraints,
we take $E_i = \infty$
and solve for the coefficients
\beq
\epd(\hat p) < \sqrt{\fr{\Fd}{\G E_f^{2d-5}L}}.
\label{bound}
\eeq
To generate a value for $L$ we consider the likely origin of the highest energy cosmic rays,
which are believed to be nuclei originating from an active galactic nucleus nearby.
For a conservative and definite number,
we take $L \approx 10$ Mpc.
Finally,
we need the final energy as the particles arrive at 
\begin{table}[H]
\begin{center}
\begin{tabular}{c|c|c|c||c|c|c|c}
$j$ & Lower bound & Coeff.\ & Upper bound & $j$ & Lower bound & Coeff.\ & Upper bound\\
\hline\hline
0	 & 						 & 	 $      \cof{  10   }{  0   0   } $ 	 & 	$<	2	\times 10^{  	-66	  }$ 	 & 	6	 & 	   $   	-2	\times 10^{	-61	  }  < $ 	 & 	 $      \cof{  10   }{  6   0   } $ 	 & 	$ <	2	\times 10^{  	-61	  }  $   	\\ \cline{1-4}
1	 & 	   $   	-1	\times 10^{	-61	  }  < $ 	 & 	 $      \cof{  10   }{  1   0   } $ 	 & 	$ <	2	\times 10^{  	-61	  }  $   	 & 	       	 & 	   $   	-1	\times 10^{	-61	  }  < $ 	 & 	 $  \re \cof{  10   }{  6   1   } $ 	 & 	$ <	1	\times 10^{  	-61	  }  $   	\\
       	 & 	   $   	-1	\times 10^{	-61	  }  < $ 	 & 	 $  \re \cof{ 10  }{  1   1   } $ 	 & 	$ <	1	\times 10^{  	-61	  }  $   	 & 	       	 & 	   $   	-1	\times 10^{	-61	  }  < $ 	 & 	 $  \im \cof{  10   }{  6   1   } $ 	 & 	$ <	1	\times 10^{  	-61	  }  $   	\\
       	 & 	   $   	-1	\times 10^{	-61	  }  < $ 	 & 	 $  \im \cof{  10   }{  1   1   } $ 	 & 	$ <	1	\times 10^{  	-61	  }  $   	 & 	       	 & 	   $   	-8	\times 10^{	-62	  }  < $ 	 & 	 $  \re \cof{  10   }{  6   2   } $ 	 & 	$ <	2	\times 10^{  	-61	  }  $   	\\ \cline{1-4}
2	 & 	   $   	-2	\times 10^{	-61	  }  < $ 	 & 	 $      \cof{  10   }{  2   0   } $ 	 & 	$ <	1	\times 10^{  	-61	  }  $   	 & 	       	 & 	   $   	-1	\times 10^{	-61	  }  < $ 	 & 	 $  \im \cof{  10   }{  6   2   } $ 	 & 	$ <	1	\times 10^{	-61	  }  $   	\\
       	 & 	   $   	-1	\times 10^{	-61	  }  < $ 	 & 	 $  \re \cof{  10   }{  2   1   } $ 	 & 	$ <	1	\times 10^{  	-61	  }  $   	 & 	       	 & 	   $   	-1	\times 10^{	-61	  }  < $ 	 & 	 $  \re \cof{  10   }{  6   3   } $ 	 & 	$ <	9	\times 10^{	-62	  }  $   	\\
       	 & 	   $   	-1	\times 10^{	-61	  }  < $ 	 & 	 $  \im \cof{  10   }{  2   1   } $ 	 & 	$ <	1	\times 10^{  	-61	  }  $   	 & 	       	 & 	   $   	-1	\times 10^{	-61	  }  < $ 	 & 	 $  \im \cof{  10   }{  6   3   } $ 	 & 	$ <	1	\times 10^{	-61	  }  $   	\\
       	 & 	   $   	-1	\times 10^{	-61	  }  < $ 	 & 	 $  \re \cof{  10   }{  2   2   } $ 	 & 	$ <	1	\times 10^{  	-61	  }  $   	 & 	       	 & 	   $   	-1	\times 10^{	-61	  }  < $ 	 & 	 $  \re \cof{  10   }{  6   4   } $ 	 & 	$ <	1	\times 10^{	-61	  }  $   	\\
       	 & 	   $   	-1	\times 10^{	-61	  }  < $ 	 & 	 $  \im \cof{  10   }{  2   2   } $ 	 & 	$ <	1	\times 10^{  	-61	  }  $   	 & 	       	 & 	   $   	-1	\times 10^{	-61	  }  < $ 	 & 	 $  \im \cof{  10   }{  6   4   } $ 	 & 	$ <	1	\times 10^{	-61	  }  $   	\\ \cline{1-4}
3	 & 	   $   	-2	\times 10^{	-61	  }  < $ 	 & 	 $      \cof{  10   }{  3   0   } $ 	 & 	$ <	2	\times 10^{  	-61	  }  $   	 & 	       	 & 	   $   	-1	\times 10^{	-61	  }  < $ 	 & 	 $  \re \cof{  10   }{  6   5   } $ 	 & 	$ <	1	\times 10^{	-61	  }  $   	\\
       	 & 	   $   	-1	\times 10^{	-61	  }  < $ 	 & 	 $  \re \cof{  10   }{  3   1   } $ 	 & 	$ <	1	\times 10^{  	-61	  }  $   	&	       	 & 	   $   	-1	\times 10^{	-61	  }  < $ 	 & 	 $  \im \cof{  10   }{  6   5   } $ 	 & 	$ <	1	\times 10^{	-61	  }  $   	\\
       	 & 	   $   	-1	\times 10^{	-61	  }  < $ 	 & 	 $  \im \cof{  10   }{  3   1   } $ 	 & 	$ <	1	\times 10^{  	-61	  }  $   	 & 	       	 & 	   $   	-1	\times 10^{	-61	  }  < $ 	 & 	 $  \re \cof{  10   }{  6   6   } $ 	 & 	$ <	1	\times 10^{	-61	  }  $   	\\
       	 & 	   $   	-1	\times 10^{	-61	  }  < $ 	 & 	 $  \re \cof{  10   }{  3   2   } $ 	 & 	$ <	1	\times 10^{  	-61	  }  $   	 & 	       	 & 	   $   	-1	\times 10^{	-61	  }  < $ 	 & 	 $  \im \cof{  10   }{  6   6   } $ 	 & 	$ <	1	\times 10^{	-61	  }  $   	\\ \cline{5-8}
       	 & 	   $   	-1	\times 10^{	-61	  }  < $ 	 & 	 $  \im \cof{  10   }{  3   2   } $ 	 & 	$ <	1	\times 10^{  	-61	  }  $   	 & 	7	 & 	   $   	-2	\times 10^{	-61	  }  < $ 	 & 	 $      \cof{  10   }{  7  0   } $ 	 & 	$ <	2	\times 10^{  	-61	  }  $   	\\
       	 & 	   $   	-1	\times 10^{	-61	  }  < $ 	 & 	 $  \re \cof{  10   }{  3   3   } $ 	 & 	$ <	1	\times 10^{  	-61	  }  $   	 & 	       	 & 	   $   	-9	\times 10^{	-62	  }  < $ 	 & 	 $  \re \cof{  10   }{ 7   1   } $ 	 & 	$ <	1	\times 10^{  	-61	  }  $   	\\
       	 & 	   $   	-1	\times 10^{	-61	  }  < $ 	 & 	 $  \im \cof{  10   }{  3   3   } $ 	 & 	$ <	1	\times 10^{  	-61	  }  $   	 & 	       	 & 	   $   	-1	\times 10^{	-61	  }  < $ 	 & 	 $  \im \cof{  10   }{ 7   1   } $ 	 & 	$ <	9	\times 10^{  	-62	  }  $   	\\ \cline{1-4}
4	 & 	   $   	-2	\times 10^{	-61	  }  < $ 	 & 	 $      \cof{  10   }{  4   0   } $ 	 & 	$ <	1	\times 10^{  	-61	  }  $   	 & 	       	 & 	   $   	-1	\times 10^{	-61	  }  < $ 	 & 	 $  \re \cof{  10   }{  7   2   } $ 	 & 	$ <	9	\times 10^{  	-62	  }  $   	\\
       	 & 	   $   	-2	\times 10^{	-61	  }  < $ 	 & 	 $  \re \cof{  10   }{  4   1   } $ 	 & 	$ <	1	\times 10^{  	-61	  }  $   	 & 	       	 & 	   $   	-1	\times 10^{	-61	  }  < $ 	 & 	 $  \im \cof{  10   }{  7  2   } $ 	 & 	$ <	1	\times 10^{	-61	  }  $   	\\
       	 & 	   $   	-1	\times 10^{	-61	  }  < $ 	 & 	 $  \im \cof{  10   }{  4   1   } $ 	 & 	$ <	1	\times 10^{  	-61	  }  $   	 & 	       	 & 	   $   	-1	\times 10^{	-61	  }  < $ 	 & 	 $  \re \cof{  10   }{  7   3   } $ 	 & 	$ <	1	\times 10^{	-61	  }  $   	\\
       	 & 	   $   	-1	\times 10^{	-61	  }  < $ 	 & 	 $  \re \cof{  10   }{  4   2   } $ 	 & 	$ <	1	\times 10^{  	-61	  }  $   	 & 	       	 & 	   $   	-1	\times 10^{	-61	  }  < $ 	 & 	 $  \im \cof{  10   }{  7   3   } $ 	 & 	$ <	1	\times 10^{	-61	  }  $   	\\
       	 & 	   $   	-1	\times 10^{	-61	  }  < $ 	 & 	 $  \im \cof{  10   }{  4   2   } $ 	 & 	$ <	1	\times 10^{	-61	  }  $   	 & 	       	 & 	   $   	-1	\times 10^{	-61	  }  < $ 	 & 	 $  \re \cof{  10   }{  7   4   } $ 	 & 	$ <	1	\times 10^{	-61	  }  $   	\\
       	 & 	   $   	-1	\times 10^{	-61	  }  < $ 	 & 	 $  \re \cof{  10   }{  4   3   } $ 	 & 	$ <	9	\times 10^{	-62	  }  $   	 & 	       	 & 	   $   	-1	\times 10^{	-61	  }  < $ 	 & 	 $  \im \cof{  10   }{  7   4   } $ 	 & 	$ <	1	\times 10^{	-61	  }  $   	\\
       	 & 	   $   	-1	\times 10^{	-61	  }  < $ 	 & 	 $  \im \cof{  10   }{  4   3   } $ 	 & 	$ <	1	\times 10^{	-61	  }  $   	 & 	       	 & 	   $   	-1	\times 10^{	-61	  }  < $ 	 & 	 $  \re \cof{  10   }{  7   5   } $ 	 & 	$ <	1	\times 10^{	-61	  }  $   	\\
       	 & 	   $   	-1	\times 10^{	-61	  }  < $ 	 & 	 $  \re \cof{  10   }{  4   4   } $ 	 & 	$ <	1	\times 10^{	-61	  }  $   	 & 	       	 & 	   $   	-1	\times 10^{	-61	  }  < $ 	 & 	 $  \im \cof{  10   }{  7   5   } $ 	 & 	$ <	1	\times 10^{	-61	  }  $   	\\
       	 & 	   $   	-1	\times 10^{	-61	  }  < $ 	 & 	 $  \im \cof{  10   }{  4   4   } $ 	 & 	$ <	1	\times 10^{	-61	  }  $   	 & 	       	 & 	   $   	-1	\times 10^{	-61	  }  < $ 	 & 	 $  \re \cof{  10   }{  7  6   } $ 	 & 	$ <	1	\times 10^{	-61	  }  $   	\\
5	 & 	   $   	-1	\times 10^{	-61	  }  < $ 	 & 	 $      \cof{  10   }{  5   0   } $ 	 & 	$ <	2	\times 10^{  	-61	  }  $   	 & 	       	 & 	   $   	-1	\times 10^{	-61	  }  < $ 	 & 	 $  \im \cof{  10   }{  7   6   } $ 	 & 	$ <	1	\times 10^{	-61	  }  $   	\\ \cline{1-4}
       	 & 	   $   	-1	\times 10^{	-61	  }  < $ 	 & 	 $  \re \cof{  10   }{  5   1   } $ 	 & 	$ <	1	\times 10^{  	-61	  }  $   	 & 	       	 & 	   $   	-1	\times 10^{	-61	  }  < $ 	 & 	 $  \re \cof{  10   }{  7   7   } $ 	 & 	$ <	1	\times 10^{	-61	  }  $   	\\
       	 & 	   $   	-1	\times 10^{	-61	  }  < $ 	 & 	 $  \im \cof{  10   }{  5   1   } $ 	 & 	$ <	1	\times 10^{  	-61	  }  $   	 & 	       	 & 	   $   	-2	\times 10^{	-61	  }  < $ 	 & 	 $  \im \cof{  10   }{  7   7   } $ 	 & 	$ <	1	\times 10^{	-61	  }  $   	\\ \cline{5-8}
       	 & 	   $   	-1	\times 10^{	-61	  }  < $ 	 & 	 $  \re \cof{  10   }{  5   2   } $ 	 & 	$ <	1	\times 10^{  	-61	  }  $   	 & 	8	 & 	   $   	-2	\times 10^{	-61	  }  < $ 	 & 	 $      \cof{  10   }{  8   0   } $ 	 & 	$ <	2	\times 10^{  	-61	  }  $   	\\
       	 & 	   $   	-1	\times 10^{	-61	  }  < $ 	 & 	 $  \im \cof{  10   }{  5   2   } $ 	 & 	$ <	1	\times 10^{	-61	  }  $   	 & 	       	 & 	   $   	-1	\times 10^{	-61	  }  < $ 	 & 	 $  \re \cof{  10   }{  8   1   } $ 	 & 	$ <	1	\times 10^{  	-61	  }  $   	\\
       	 & 	   $   	-1	\times 10^{	-61	  }  < $ 	 & 	 $  \re \cof{  10   }{  5   3   } $ 	 & 	$ <	2	\times 10^{	-61	  }  $   	 & 	       	 & 	   $   	-1	\times 10^{	-61	  }  < $ 	 & 	 $  \im \cof{  10   }{  8   1   } $ 	 & 	$ <	1	\times 10^{  	-61	  }  $   	\\
       	 & 	   $   	-1	\times 10^{	-61	  }  < $ 	 & 	 $  \im \cof{  10   }{  5   3   } $ 	 & 	$ <	1	\times 10^{	-61	  }  $   	 & 	       	 & 	   $   	-1	\times 10^{	-61	  }  < $ 	 & 	 $  \re \cof{  10   }{  8   2   } $ 	 & 	$ <	1	\times 10^{  	-61	  }  $   	\\
       	 & 	   $   	-1	\times 10^{	-61	  }  < $ 	 & 	 $  \re \cof{  10   }{  5   4   } $ 	 & 	$ <	1	\times 10^{	-61	  }  $   	 & 	       	 & 	   $   	-1	\times 10^{	-61	  }  < $ 	 & 	 $  \im \cof{  10   }{  8   2   } $ 	 & 	$ <	1	\times 10^{	-61	  }  $   	\\
       	 & 	   $   	-1	\times 10^{	-61	  }  < $ 	 & 	 $  \im \cof{  10   }{  5   4   } $ 	 & 	$ <	1	\times 10^{	-61	  }  $   	 & 	       	 & 	   $   	-1	\times 10^{	-61	  }  < $ 	 & 	 $  \re \cof{  10   }{  8   3   } $ 	 & 	$ <	1	\times 10^{	-61	  }  $   	\\
       	 & 	   $   	-1	\times 10^{	-61	  }  < $ 	 & 	 $  \re \cof{  10   }{  5   5   } $ 	 & 	$ <	9	\times 10^{	-62	  }  $   	 & 	       	 & 	   $   	-1	\times 10^{	-61	  }  < $ 	 & 	 $  \im \cof{  10   }{  8   3   } $ 	 & 	$ <	1	\times 10^{	-61	  }  $   	\\
       	 & 	   $   	-1	\times 10^{	-61	  }  < $ 	 & 	 $  \im \cof{  10   }{  5   5   } $ 	 & 	$ <	1	\times 10^{	-61	  }  $   	 & 	       	 & 	   $   	-1	\times 10^{	-61	  }  < $ 	 & 	 $  \re \cof{  10   }{  8   4   } $ 	 & 	$ <	2	\times 10^{	-61	  }  $   	\\
	 & 						 & 		 & 						 & 	       	 & 	   $   	-1	\times 10^{	-61	  }  < $ 	 & 	 $  \im \cof{  10   }{  8   4   } $ 	 & 	$ <	1	\times 10^{	-61	  }  $   	\\
	 & 						 & 		 & 						 & 	       	 & 	   $   	-1	\times 10^{	-61	  }  < $ 	 & 	 $  \re \cof{  10   }{  8   5   } $ 	 & 	$ <	1	\times 10^{	-61	  }  $   	\\
	 & 						 & 		 & 						 & 	       	 & 	   $   	-1	\times 10^{	-61	  }  < $ 	 & 	 $  \im \cof{  10   }{  8   5   } $ 	 & 	$ <	1	\times 10^{	-61	  }  $   	\\
	 & 						 & 		 & 						 & 	       	 & 	   $   	-1	\times 10^{	-61	  }  < $ 	 & 	 $  \re \cof{  10   }{  8   6   } $ 	 & 	$ <	2	\times 10^{	-61	  }  $   	\\
	 & 						 & 		 & 						 & 	       	 & 	   $   	-1	\times 10^{	-61	  }  < $ 	 & 	 $  \im \cof{  10   }{  8  6   } $ 	 & 	$ <	1	\times 10^{	-61	  }  $   	\\
	 & 						 & 		 & 						 & 	       	 & 	   $   	-1	\times 10^{	-61	  }  < $ 	 & 	 $  \re \cof{  10   }{  8   7   } $ 	 & 	$ <	1	\times 10^{	-61	  }  $   	\\
	 & 						 & 		 & 						 & 	       	 & 	   $   	-1	\times 10^{	-61	  }  < $ 	 & 	 $  \im \cof{  10   }{  8   7  } $ 	 & 	$ <	1	\times 10^{	-61	  }  $   	\\
	 & 						 & 		 & 						 & 	       	 & 	   $   	-1	\times 10^{	-61	  }  < $ 	 & 	 $  \re \cof{  10   }{  8   8   } $ 	 & 	$ <	1	\times 10^{	-61	  }  $   	\\
	 & 						 & 		 & 						 & 	       	 & 	   $   	-1	\times 10^{	-61	  }  < $ 	 & 	 $  \im \cof{  10   }{  8   8   } $ 	 & 	$ <	1	\times 10^{	-61	  }  $   	\\	
\end{tabular}
\caption{\label{d8}
Conservative constraints on coefficients $\cof{10}{jm}$ in GeV$^{-6}$.}
\end{center}
\end{table}
\noindent
Earth,
which is based on the observed energy.
Again proceeding toward conservative constraints,
we assume that a partonic fermion carrying 10\% of the proton energy
in an iron nucleus is the radiator of gravitons.
Hence we take $E_f$ as 1/560 of the observed energy.

One observation typically yields a one-sided constraint on a combination of coefficients for Lorentz violation,
since gravitational \v Cerenkov radiation is only possible for particles moving faster than the speed of gravity.
However,
with the exception of the isotropic coefficients,
2-sided constraints are achieved using cosmic rays originating from multiple places on the sky.
In Ref.\ \cite{cer},
a series of 6 models were constrained.
The models included 3 isotropic models containing the isotropic coefficient at $d=4$, $d=6$, and $d=8$ respectively,
and 3 anisotropic models containing the rest of the coefficients at each dimension.
The numerical constraints were obtained from the energy and direction of origin for observed cosmic rays
using a linear programming scheme (for details, see \cite{diaz2}).
Here we perform the same operations on the same data
for 2 more models:
an isotropic $d=10$ model and an anisotropic $d=10$ model.
The constraints that result,
which are the first on $d=10$ coefficients in the gravity sector,
appear in Table 3.

\section{summary}

This work summarizes and expands upon gravity-related tests
of Lorentz symmetry in the SME.
We begin in Sec.\ \ref{basics} by providing some examples of symmetry and symmetry violation
that highlight CPT and Lorentz violation conceptually
and provide context for the SME structure reviewed in Sec.\ \ref{SME}.
The gravitationally coupled fermion sector
and the pure gravity sector are presented as generic examples of the SME construction,
and to provided context for the phenomenology.
Existing and proposed tests are summarized in Sec.\ \ref{grev}.
Section \ref{cer} provides some more detail on one recent test
that achieved tight first constraints on dimension 6 and 8 coefficients in the gravity sector
through the analysis of gravitational \v Cerenkov radiation effects on cosmic rays.
Following review of that work,
81 tight new constraints on dimension 10 coefficients are achieved following
the same methods.
Both the new results achieved and the existing proposals for expansion
discussed here highlight the bright future
for the continued expansion of tests of Lorentz invariance
in gravitational experiments.

\bibliographystyle{mdpi}

\renewcommand\bibname{References}



\end{document}